\documentclass{article} 
\usepackage{iclr2019_conference,times}
\usepackage{graphicx} 
\graphicspath{ {./images/} }

\usepackage{amsmath,amsfonts,bm}

\def\eqref#1{equation~\ref{#1}}

\def\1{\bm{1}}

\DeclareMathAlphabet{\mathsfit}{\encodingdefault}{\sfdefault}{m}{sl}
\SetMathAlphabet{\mathsfit}{bold}{\encodingdefault}{\sfdefault}{bx}{n}

\usepackage{hyperref}
\usepackage{url}
\usepackage{wrapfig}

\title{ProDyn0: Inferring calponin homology domain stretching behavior using graph neural networks}

 \author{Ali Madani, Cyna Shirazinejad, Jia Rui Ong, Hengameh Shams, Mohammad Mofrad \\
 Molecular Cell Biomechanics Laboratory\\
 University of California, Berkeley\\
 Berkeley, CA 94720, USA \\
 \texttt{\{madani,mofrad\}@berkeley.edu} 
 }

\iclrfinalcopy 
\begin{document}

\maketitle\

\begin{abstract}

Graph neural networks are a quickly emerging field for non-Euclidean data that leverage the inherent graphical structure to predict node, edge, and global-level properties of a system. Protein properties can not easily be understood as a simple sum of their parts (i.e. amino acids), therefore, understanding their dynamical properties in the context of graphs is attractive for revealing how perturbations to their structure can affect their global function. To tackle this problem, we generate a database of 2020 mutated calponin homology (CH) domains undergoing large-scale separation in molecular dynamics. To predict the mechanosensitive force response, we develop neural message passing networks and residual gated graph convnets which predict the protein dependent force separation at 86.63 percent, 81.59 $kJmol^{-1}nm^{-1}$ MAE, 76.99 $psec$ MAE for force mode classification, max force magnitude, max force time respectively-- significantly better than non-graph-based deep learning techniques. Towards uniting geometric learning techniques and biophysical observables, we premiere our simulation database as a benchmark dataset for further development/evaluation of graph neural network architectures.\footnote{\url{https://github.com/a-mad/prodyn}}

\end{abstract}

\section{Introduction}

The primary sequence of a well-ordered protein encodes its structural information as well as its functional properties. Proteins undergo dynamic shifts in conformation to perform their designated tasks. The growth of computational resources and simulation algorithms allow for studying molecular trajectories at atomic resolution commonplace. However, because of the heterogeneity of bio-molecules and their chemical environments, computational models can suffer in performance when tasked with predicting molecular conformations, physical and chemical properties, and protein-protein interaction networks. Along with the growing excitement in machine learning in tangent fields, there is emerging interest in data-driven approaches to studying biological function at scales ranging from biochemical properties to tissues mechanics (\cite{glazer2008combining}; \cite{PhysRevB.92.094306}; \cite{schutt2017quantum}; \cite{gastegger2017machine}; \cite{madani2018deep}; \cite{wu2018moleculenet}; \cite{ramakrishnan2015electronic}).

Mechanosensitive proteins are central to a plethora of biological phenomena (\cite{galkin2010opening}). Our goal in this paper is to bridge the primary sequence of proteins to their intrinsic force observables while undergoing large-scale conformational changes. To accomplish this, we generated a molecular dynamics database of trajectories of two calponin homology domain mutants being pulling apart [Figure 1] by Steered Molecular Dynamics (SMD) (\cite{isralewitz2001steered}). SMD is a technique that allows for biasing the interaction potential between explicit regions in a simulation; in our case, two globular domains, CH1 and CH2, are gradually pulled apart during the simulation. While these simulations are computationally expensive and require careful consideration for proper physical dynamics, they provide information at the atomic resolutions about the sensitivity of the roles mutations play in response to mechanical loads in proteins. Our aim to cast protein structures as graphical models allows us to take the first imperative steps towards building relationships between sequences and physical observables that can be computed with more expensive techniques such as molecular dynamics. We show that the force between the two CH domains, a global property over a long simulated trajectory, can accurately be predicted using graph neural networks.

\begin{figure}[h]
\caption{Workflow Schematic: From Biophysics to Graph Neural Networks}
\centering
\includegraphics[width=1.0\textwidth]{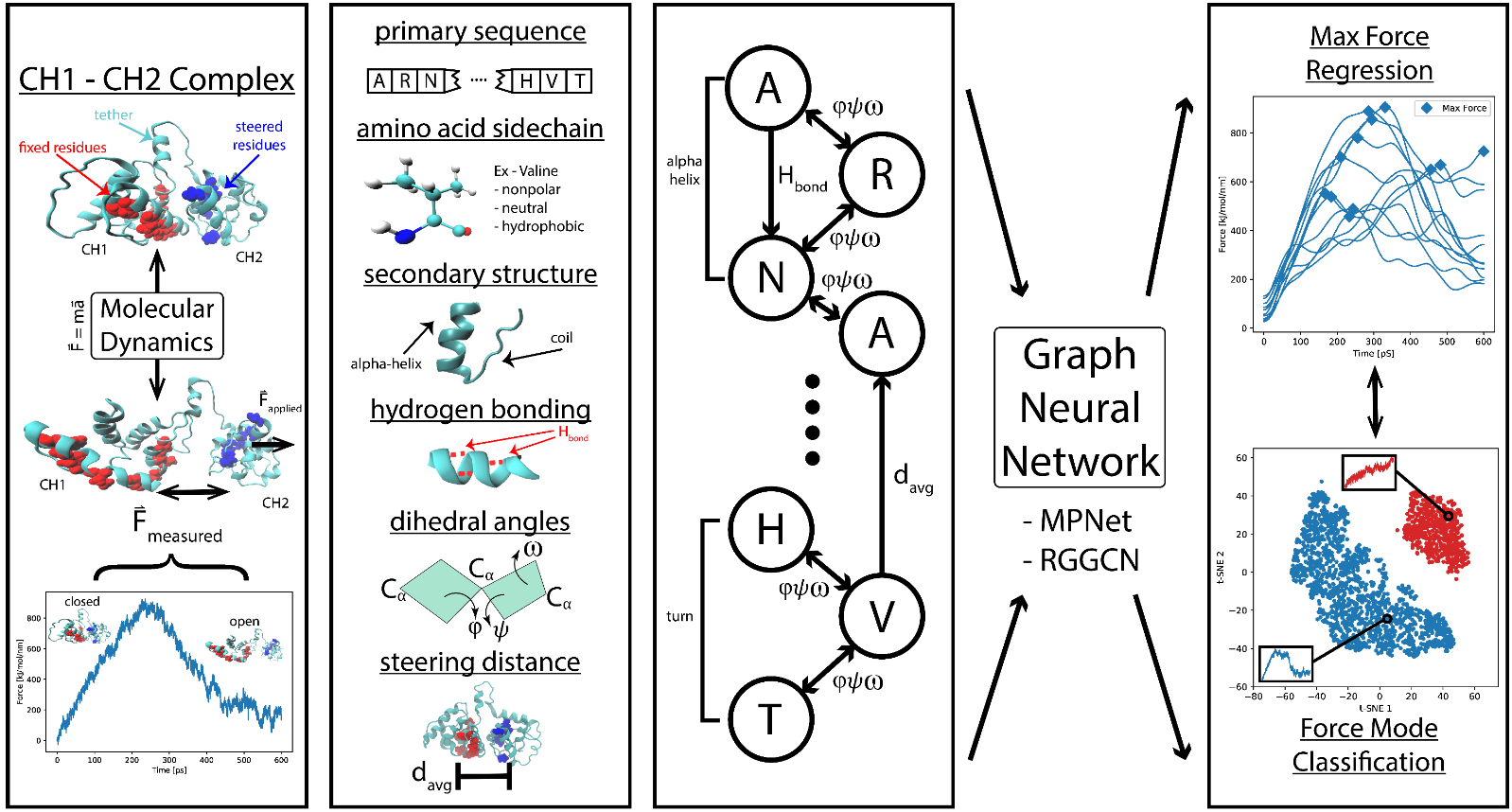}
\end{figure}

\section{Methods}

\subsection{Generating a database of point-mutated protein dynamics}

To generate models to predict protein dynamics, we established the first ProDyn dataset with high-throughput \textit{in silico} mutagenesis experiments of proteins undergoing large-scale conformational changes. The generated trajectories span 2020 unique point-mutations along two calponin homology domains, CH1 and CH2, connected by a tether that comprise a total of 224 residues. A full description of the molecular dynamics methods can be found in Supplementary Materials. The data per simulation was processed to yield a graphical input representation of residue interactions in terms of node and edge features, in addition to an output representation for force profile classification and regression. We then construct and compare conventional and graph neural network architectures to predict the force characteristics of our biophysical system.

\subsection{Data Pre-processing}
A molecular dynamics (MD) simulation under a given force field outputs a trajectory file that includes type, position, and velocity information for every atom in the system for each time step. There are several prediction problems of interest that can be encapsulated by our data. In this study, our specific objective is to predict the force characteristics for a given protein initial state (i.e. residue information and interactions pre-steering).

The input representation is a directed graph, $G$, with node features, $x_v$ and edge features, $e_{vw}$. As shown in Table 1, node features comprise of the residue type, residue properties, and initial secondary structure type as calculated by mdtraj (\cite{McGibbon2015MDTraj}). Edge features currently consist of pair-wise dihedral angles, Kabsch-Sander (K-S) hydrogen bonds (\cite{kabsch1983dictionary}), and the initial distance between steered and fixed residues. 

The output representation, a graph-level target of protein mechanosensitivity, is of two types: regression and classification. Each SMD simulation exhibits a characteristic force response over time of the simulation. The regression target is a 2 dimensional vector encoding maximum pulling force magnitude and maximum pulling force time-point respectively. The classification target is a one-hot encoded 2-dimensional vector that encodes a force mode category. The force mode categories were ascertained via spectral k-means (k=2) clustering on the force-response graphs in a lower dimensional t-SNE (\cite{maaten2008visualizing}) derived space as shown in Figure 2. There emerged two clearly well-separated clusters which we utilized as categories for graph-level classification. Lastly, all continous data was scaled to [0,1], and the entire dataset was split randomly into a 404 sample held-out test set and 1616 sample 80-20 training/validation split. For the classification task, we compute the accuracy and macro-averaged F1-score as performance metrics. For the regression task, we calculate the mean absolute error (MAE) for maximum force magnitude and maximum force time-point.

\begin{center}
\vspace{-1.0mm}
\begin{tabular}{ |c|c|c|c|c| } 
\multicolumn{5}{c}{Table 1 - Input Representation Features} \\
\hline
Feature & Node/Edge & Type & Index & Examples \\
\hline

Residue Type & Node & One-hot & $[0:20]$ & ALA, GSP, ALY, ... \\

Residue Secondary Structure & Node & One-hot & $[21:26]$ & alpha-helix, turn, ... \\

Residue Properties & Node & Multi-label & $[27:43]$ & acidic, polar, ... \\

Dihedral-$\phi$ & Edge & Continuous & $[0]$ & $0-2\pi$ \\
Dihedral-$\psi$ & Edge & Continuous & $[1]$ & $0-2\pi$ \\
Dihedral-$\omega$ & Edge & Continuous & $[2]$ & $0-2\pi$ \\
Hydrogen Bond & Edge & Continuous & $[3]$ & K-S energy \\
Steer-to-Fixed Residue Dist & Edge & Continuous & $[4]$ & average distance\\

\hline
\end{tabular}
\end{center}

\subsection{Baseline Conventional Deep Learning Models}
As a comparison to graph-based neural networks described below, conventional multi-layer perceptron (MLP) and gated recurrent unit (GRU) were trained (\cite{cho2014learning}). The graph node features $x_v$ were provided as input either by concatenation or sequentially. For the MLP architectures with one vector input of concatenated features, the number of layers [1-5], activation functions [ReLU, eLU], nodes per layer [128-512], dropout regularization [0.0-0.5], and loss functions [L1, L2, logcosh, cross-entropy] were experimented with. For the GRU+MLP architecture with inputs features passed sequentially through the GRU, we experimented with different GRU implementations- varying hidden dimensions [16-64], stacking [1-2], bi-directionality, regularization [via dropout], and outputs [last layer depth-wise vs time-wise].

\subsection{Neural Message Passing Network Model}
In line with \cite{gilmer2017neural}, we developed various neural message passing network (MPNet) architectures. Our models learn to compute a function of the entire graph by learning features, invariant to graph isomorphisms, through a message passing algorithm. We define the hidden state of each node in the graph by $h^t_v$ with $h^0_v = x_v$ where v denotes residue/node number and t denotes the cycle number, and the directed edge from node w to v as $e_{v\leftarrow w}$. The message passing scheme runs for a total of $T$ cycles which can be interpreted as the message-passing neighborhood size. During each cycle, messages are passed between a neighboring node and the current node based on the edge direction. There is a unique message function, $M$, that computes the message given the current node features, neighboring node features, and edge features which is then aggregated into one message update per node, $m_v^t$. This message is then used to update the hidden node state by the function, $U$. The final hidden states of all nodes are then passed through a readout function, $R$, which yields an output layer of the dimension of the sample target. To formalize, for each message passing cycle the node states are updated in the following fashion:
\begin{equation}
    m_v^t = \frac{1}{|N(v)|}\sum_{w\in N(v)}M(h_v^t,h_w^t,e_{v\leftarrow w})
\end{equation}
\begin{equation}
    h_v^{t+1} = U(h_v^t,m_v^t) = h_v^t + m_v^t
\end{equation}
where $N(v)$ denotes the neighbors of v in graph G. The message function, M, is a 3-layer MLP with 0.4 dropout, 256 hidden layer dimensions, and output layer of the same dimension as node features. After T cycles, the readout function R is applied as follows:
\begin{equation}
    \hat{y} = R(\{h_v^T | v\in G\})
\end{equation}
We did not see a significant increase in performance for increasing T cycles and present our MP models for T=1 in Results. For the choice of R, we experimented with a max-pooling operation over the feature dimension and also feeding the nodes in primary sequence order to a GRU. The final output time-wise of the GRU is fed to a 2-layer MLP with 0.5 dropout. The final loss function, $L(\hat{y},y)$ was either a L1-norm for regression or softmax+cross-entropy for classification. Models were trained using stochastic gradient descent with ADAM (\cite{kingma2014adam}) and batch size of 25.

\subsection{Residual Gated Graph ConvNet Model}
Residual Gated Graph ConvNets (RGGCN) is a method introduced by \cite{bresson2017residual} that combines the vanilla graph ConvNet (\cite{sukhbaatar2016learning}) and the edge gating mechanism (\cite{marcheggiani2017encoding}) with residual connections for problems involving variable graphs. We extend this method by incorporating edge features, such that each graph convolution layer follows the layer-wise propagation rule:
$$
h _ { i } ^ { \ell + 1 } = \operatorname { ReLU } \left( U ^ { \ell } h _ { i } ^ { \ell } + \sum _ { j \rightarrow i } \eta _ { j \rightarrow i  } \odot ( V ^ { \ell } h _ { j } ^ { \ell } +  D ^ { \ell } e _ { j \rightarrow i } ) \right)
$$
where $h _ { i } ^  { \ell }$ denote features of node  $i$ at layer $l$, $e _ { j \rightarrow i  }$  denote features from edge $ j \rightarrow i$, and edge gates $\eta _ { j \rightarrow i  } = \sigma \left( A ^ { \ell } h _ { i } ^ { \ell } + B ^ { \ell } h _ { j } ^ { \ell } + C ^ { \ell } e _ { j \rightarrow i  } \right)$. $U$, $V$, $A$, $B$, $C$ and $D$ are learnable parameters.

In our experiments, we used a 4-layer RGGCN and applied max pooling as our readout function. For classification, our model had 50 hidden dimensions and a single linear layer. 
For regression, our model had 256 hidden dimensions with a 2-layer MLP. Batch normalization was employed, as with residual connections between successive convolution layers.

Our models were initialized using Glorot initialization (\cite{glorot2010understanding}). We used the Adam SGD optimizer with an initial learning rate of 0.0005, and computed loss using binary cross-entropy for classification, and L1 for regression. Dropout rate was set to 0.4 for both node features and edge gates in classification, and only for the 2-layer MLP in regression.

\section{Results and Discussion}
One of our main achievements is creating a database that is uniquely well-positioned to benchmark the rapid development and evaluation of novel graph neural network architectures on both a graph-level classification and regression task. As shown in Table 2, our baseline non-graph-based deep learning models-- both (1) variations of concatenating features and processing through an MLP and (2) sequentially feeding primary sequence-ordered nodes through a GRU -- were unable to learn effectively and predicted tightly near the majority/mean target. Within our data size scale, the complexity of many biophysical phenomena, through initial states and short/long-range bonded and non-bonded interactions partially captured in this study, is too difficult to train effectively unconstrained via an MLP or recurrently via a GRU.

\begin{center}
\begin{tabular}{ |c|c|c|c|c| } 
\multicolumn{5}{c}{Table 2 - Prediction Performance of Various Architectures} \\
\hline
Model & Accuracy & F1-Score & $\text{Force}_{mag}$ MAE & $\text{Force}_{time}$ MAE \\
\hline
Conventional & 70.54 & 0.414 & 89.68 & 111.39 \\
MPNet + Maxpool  & 77.09 & 0.673 & 85.39 & 92.44 \\
MPNet + GRU & \textbf{86.63} & 0.839 & \textbf{81.59} & \textbf{76.99} \\
RGGCN + Maxpool & \textbf{86.39} & \textbf{0.904} & 89.00 & 100.52 \\
\hline

\end{tabular}
\end{center}

Our goal was to experiment with 2 types of graphical neural networks: our own designed MPNet and a state-of-the-art graph technique such as the RGGCN. Both graph neural network architectures, MPNet and RGGCN, capture the graphical structure, information communication, and structural invariances to learn an effective node embedding for classification/regression tasks. In end-to-end training after node embedding, a readout function is applied that is agnostic to the number of nodes/residues. Among functions invariant to graph isomorphism, we observed that maxpool outperformed average pool. As proteins have multiple hierarchies of structure, we can view the MPNet with GRU readout as encoding the protein structure hierarchy with the GRU applied after message passing cycle, as to allow message passing interactions to be communicated between nodes before the recurrent primary structure network. The choice of GRU readout is likely more effective for protein-like graph systems; the RGGCN+Maxpool would be more generalizable otherwise.

\begin{figure}[h]
    \includegraphics[width=\textwidth]{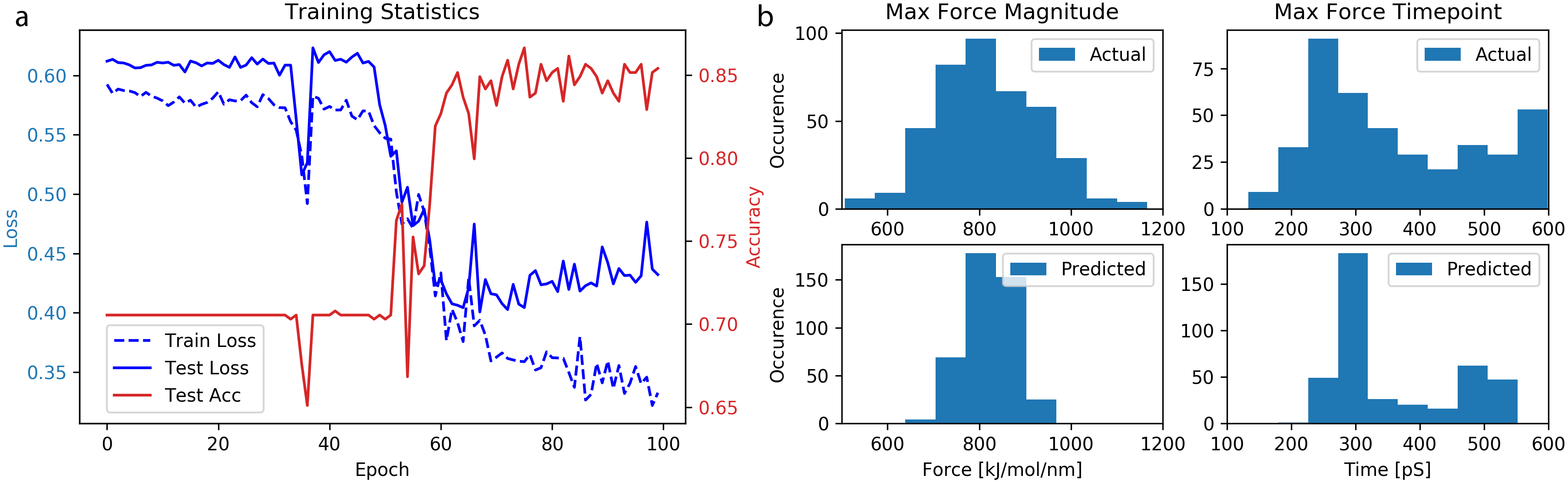}
  \caption{MPNet + GRU Performance. a) Classification training statistics b) Regression histograms of predicted vs actual values}
\end{figure}

As shown in Table 2, all graph neural networks significantly outperform conventional deep learning techniques. The RGGCN+Maxpool and MPNet+GRU perform well on the classification task at around ~86\%. For the regression task, the MPNet+GRU is able to learn with a MAE of 81.59 $kJmol^{-1}nm^{-1}$ MAE on max force magnitude prediction and 76.99 $psec$ MAE on max force time prediction. To put into context the performance to the natural stochasticity of the MD system, we ran duplicate SMD simulations and observed the mean standard deviation of the max force mag was 57.94. Also, in Figure 2, we show the well-behaved convergence of the classification model training in addition to the resemblant output statistics of the true and predicted targets for regression.

Future work can include learning edge states in addition to node states, set2vec as readout function (\cite{vinyals2015order}) or GRU as an update function, and experimenting with unsupervised graph embedding techniques such as \cite{velivckovic2018deep}. From the biophysical perspective, we can expand the choice of node/edge features, try different protein systems, or formulate a plethora of pertinent biophysical prediction tasks.

To conclude, the progress of geometric deep learning techniques is inextricably linked to the quality of benchmark datasets-- analagous to the impact of benchmark datasets for vision and language. We view our work as laying the groundwork for the intersection of graph neural network and biophysics researchers for the advancement of both fields.

\newpage
\bibliography{iclr2019_conference}
\bibliographystyle{iclr2019_conference}

\newpage 
\section{Supplementary Materials}

The creation of the mutation library was primarily informed by a BLOSUM95 matrix displaying frequencies of swaps between pairs of residues in closely-related protein homologues (\cite{henikoff1992amino}). In addition, an alanine scan over most of the sequence was made to complement work done by saturation mutagenesis experiments (\cite{weiss2000rapid}). To augment our data set beyond this, a random sample of positions were mutated to random residues with no further bias. 

To generate the database of single point-mutated calponin homology domain trajectories, all-atom molecular dynamics was performed. Experimental procedures were largely adapted from previous work documenting the SMD experimental protocol for the wild-type CH1-CH2 domain (\cite{shams2016dynamic}). The wild type structure from Shams et al. was modified at single residue locations using a MODELLER (https://salilab.org/modeller/) script that swaps residues, optimizes dihedral angles, reduces bad contacts with conjugate gradient steps, and performs a few steps of molecular dynamics to allow the swapped residue to relax (\cite{vsali1993comparative}). 

All subsequent molecular dynamics was performed with GROMACS (\cite{van2005gromacs}; \cite{lindahl2001gromacs}; \cite{berendsen1995gromacs}) using the CHARMM27 all-atom force-field. Pulling simulations were performed in a 11.5 x 11.5 x 13 nm3 box solvated with SPC216 water molecules and 100 mM KCL. The system was minimized with steepest-descent to remove bad contacts until the total energy was less than 1000.0 kJ/mol/nm. Periodic boundary conditions were set in all 3 directions, a Verlet cutoff scheme truncated non-bonded long-range interactions at 1.2 nm, and PME electrostatics with Fourier spacing of 0.12 nm was used. Following minimization, each system was equilibrated for 100 ps in NVT followed by 100 ps of NPT, both using 2 fs time steps and a target temperature of 300K and 1 bar, respectively. The equilibration was closely monitored for each mutation to ensure that the system was properly sampling the correct equilibrium distribution before steering experiments were performed.

SMD experiments were performed by anchoring down and pulling on a selected group of heavy carbon atoms on the interface between the CH1 and CH2. To maintain this experimental control, residues that include these atoms were not mutated in our experiments. All other parameters of our simulations is held constant except for a single point mutation that is unique to each simulation.  The reaction coordinate distance is defined here as the distance between the center of masses of the fixed and steered residues. Steering was performed by moving an umbrella potential with stiffness 1255 kJ mol-1 nm-2 at a rate of 2 nm/ps along a reaction coordinate defined by the axis connecting the center of masses of the fixed and steered atoms. Temperature coupling was performed by a Nose-Hoover thermostat targeting 310K and coupling was maintained at 1 bar with an isotropic Parrinello-Rahman piston for the duration of the pulling experiment. Throughout the 600 ps pulling experiment, readouts of the reaction coordinate distance and the force between the fixed and steered atoms were measured. The maximum force applied between the domains as well as the entirety of the force profile are the observables of interest predicted with our neural networks. Figure 1 shows the force profiles can be vary dramatically in shape for different mutations, while remaining relatively stationary for different initial conditions per mutation. 

\end{document}